\documentclass[conference]{IEEEtran}
\IEEEoverridecommandlockouts
\usepackage{cite}
\usepackage{amsmath,amssymb,amsfonts}
\usepackage{algorithmic}
\usepackage{graphicx}
\usepackage{makecell}
\usepackage{multirow}
\usepackage{bm}
\usepackage{subcaption}
\usepackage{textcomp}
\usepackage{cleveref}
\usepackage{acronym}
\usepackage{dsfont}
\usepackage{colortbl}
\usepackage[table]{xcolor}
\usepackage{todonotes}
\usepackage{xcolor}
\def\BibTeX{{\rm B\kern-.05em{\sc i\kern-.025em b}\kern-.08em
    T\kern-.1667em\lower.7ex\hbox{E}\kern-.125emX}}

\acrodef{MFCC}[MFCC]{Mel-Frequency Cepstral Coefficients}
\acrodef{SED}[SED]{Sound Event Detection}
\acrodef{FFT}[FFT]{Fast Fourrier Transform}
\acrodef{GMM}[GMM]{Gaussian Mixture Model}
\acrodef{SVM}[SVM]{Support Vector Machine}
\acrodef{PCA}[PCA]{Principal Component Analysis}
\acrodef{CNN}[CNN]{Convolutional Neural Network}
 
\begin{document}

\title{Sound-Based Spin Estimation in Table Tennis: Dataset and Real-Time Classification Pipeline
\thanks{This work was funded by Sony AI}
}

\author{\IEEEauthorblockN{Thomas Gossard$^*$,
Julian Schmalzl$^*$,\thanks{$^*$Equal contribution}
Andreas Ziegler,
Andreas Zell
}
\IEEEauthorblockA{Cognitive Systems Group, University of Tuebingen, Germany}
}

% \and
% \and
% \and
% \and
% \and

\maketitle

\begin{abstract}
Sound can complement vision in ball sports by providing subtle cues about contact dynamics.
In table tennis, the brief, high-frequency sounds produced during racket-ball impacts carry information about the racket type, the surface contacted, and whether spin was applied.
We address three key problems in this domain: (1) precise bounce detection with millisecond-level temporal accuracy, (2) classification of bounce surface (e.g., racket, table, floor), and (3) spin detection from audio alone.
To this end, we propose a real-time-capable pipeline that combines energy-based peak detection with convolutional neural networks trained on a novel dataset of 3,396 bounce samples recorded across 10 racket configurations.
The system achieves accurate and low-latency detection of bounces, and reliably classifies both the surface of contact and whether spin was applied.
This audio-based approach opens up new possibilities for spin estimation in robotic systems and for real-time feedback in coaching tools.
We publicly release both the dataset and code to support further research.
\end{abstract}


\section{Introduction}

Although ball-based sports primarily rely on vision, auditory cues can also provide a competitive advantage.
First, the volume of the bounce sound conveys information about the force of impact. 
Cañal-Bruland et al. demonstrated that the intensity of sound can affect the anticipated trajectory of the ball in tennis~\cite{canal-bruland2018}.
However, subsequent research indicates that this effect is also context-dependent, e.g. player positions, ball trajectory before being hit~\cite{canal-bruland2022}.
Similarly, the ability to recognize smashes in volleyball or power shots in football early on can be enhanced by auditory cues~\cite{sors2017}.
The same can be said for table tennis where the impact's volume is correlated with the outgoing speed of the ball.

The usefulness of sound for table tennis was analyzed by having people play without any sound feedback.
A decrease in performance was observed when tennis players were wearing earplugs~\cite{takeuchi1993}.
Fujita et al. showed more specifically that auditory information improves response time and counterattack performance~\cite{fujita2023a}.
In table tennis, sound can provide valuable insights, such as the magnitude of spin.
Incorporating auditory data has been shown to improve the prediction of ball spin type in table tennis~\cite{park2016}.
Peterossi et al. reach a similar conclusion that the magnitude of the spin can be inferred from the sound but argue that playing experience has little influence on the level of spin prediction from sound~\cite{peterossi2017}.
In addition, one can identify the rubber used to hit the ball. 
Before table tennis rubbers were regulated to one side be red and the other black, players could have two different rubbers of the same color (e.g. sticky rubber and anti-topspin). 
As such, players would use the auditory cue to distinguish which rubber was used to serve the ball and this informed them whether spin was applied.
Stomping would conceal that sound, which would make predicting the spin more difficult.
Stomping is still observed in some players but serves other purposes such as returning faster to the ready position or out of habit.
However, there is a limit to what information can be extracted from sound.
One cannot judge the quality of a racket stroke from only the sound\cite{petri2020}. 
Moreover, while the amount of spin can be determined, predicting the direction of the spin is considerably more difficult and requires visual input~\cite{peterossi2017}.

\begin{figure}[t]
    \centering
    \includegraphics[width=0.9\linewidth]{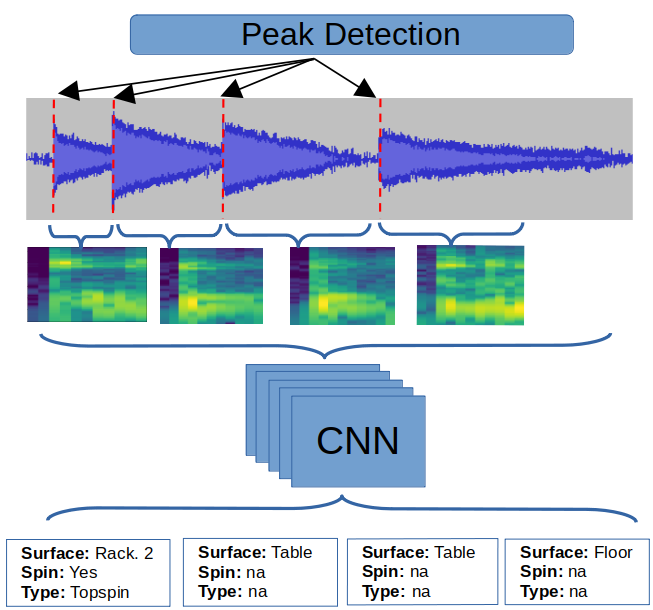}
    \caption{Overview of the proposed real-time pipeline for racket-ball bounce sound classification. Bounce candidates are detected using energy peaks based on an exponential moving average on the high-pass filtered signal. 
Detected segments are converted to Mel spectrograms and classified using CNNs to identify the bounce surface and detect whether spin was applied. 
The system achieves millisecond-level onset accuracy.}
    \label{fig:pipeline}
\end{figure}

In this paper, we introduce a real-time pipeline that tackles three core tasks in the analysis of table tennis bounce sounds: (1) millisecond-accurate bounce detection, (2) classification of the contact surface (racket, table, floor), and (3) detection of spin application from audio alone.
To support this, we collected and annotated a high-quality dataset of 3,396 bounce recordings across 10 diverse racket configurations.
This pipeline opens up promising applications in robotic spin estimation, real-time feedback systems, and audio-based coaching tools.
The dataset and code are publicly available at: \textit{https://github.com/cogsys-tuebingen/tt\_sounds}.

\section{Related work}
The ball's spin can be estimated from different information sources.
Spin can be estimated through the curve in the ball's trajectory due to the Magnus effect~\cite{wang2023,su2013,tebbe2020a,chen2010a}.
The change in direction after the bounce can also be leveraged~\cite{wang2023}.
However, what most players do is estimate the spin applied on the ball from the stroke motion~\cite{sato2020,kulkarni2021,gao2021}.
The most accurate way to estimate spin is to directly observe the ball logo~\cite{tebbe2020a, gossard2024} or a pattern drawn on the ball~\cite{gossard2023}.
However, no one has yet tried to quantify spin from audio cues in table tennis, as far as we know.

The detection and classification of the racket-ball bounce sound falls in the field of \ac{SED}.
\ac{SED} consists of detecting the temporal onset and offset of different sound events and classifying them.
\ac{SED} offers a lot of potential for automatic annotation pipelines in sports.
It can be used to detect highlights in baseball, golf, or soccer~\cite{xiong2003, baillie2003}.
It can also improve the annotation of tennis games, such as detecting when the ball is hit.

\begin{figure}
    \centering
    \begin{subfigure}[b]{0.9\linewidth}
        \includegraphics[width=\linewidth]{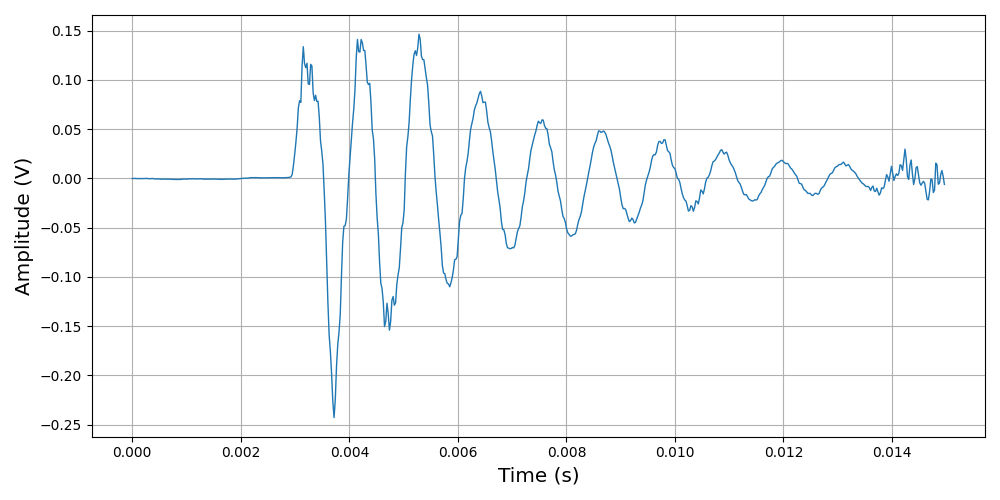}
        \caption{Waveform}
        \label{fig:waveform}
    \end{subfigure}
    \begin{subfigure}[b]{0.9\linewidth}
        \includegraphics[width=\linewidth]{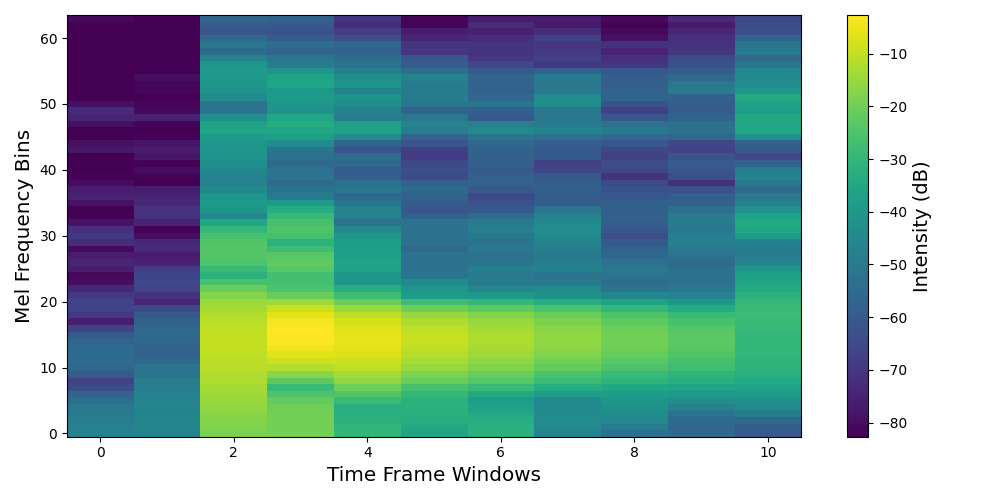}
        \caption{Mel-Spectogram}
        \label{fig:mel_spectogram}
    \end{subfigure}
    \begin{subfigure}[b]{0.9\linewidth}
        \includegraphics[width=\linewidth]{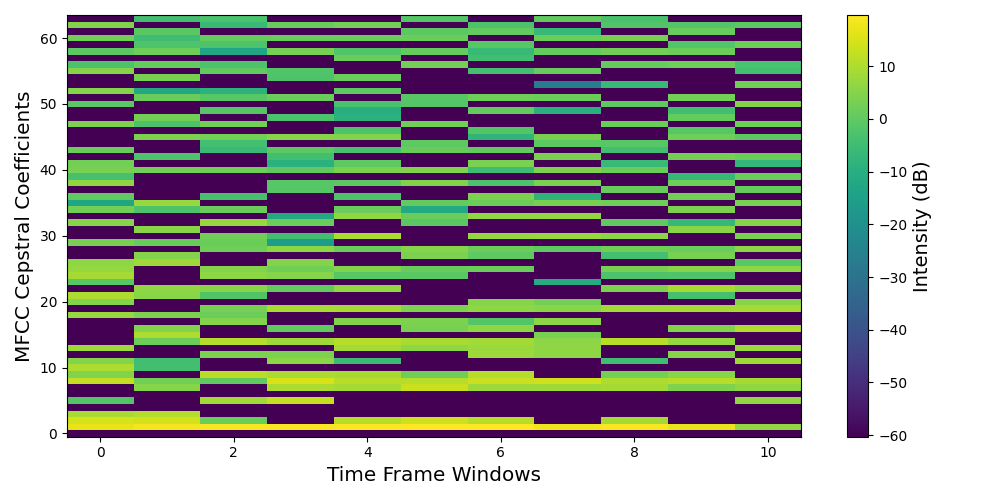}
        \caption{MFCC}
        \label{fig:mfcc}
    \end{subfigure}
    \caption{Different possible representations of the bounce sound}
    \label{fig:bounce_sound}
\end{figure}

Sound can be represented in various forms to facilitate analysis and classification.
The raw waveform (\Cref{fig:waveform}) provides a time-domain view of the signal, capturing amplitude variations but lacking explicit frequency information.
To better capture the perceptual characteristics of sound, time–frequency representations such as the Mel spectrogram (\Cref{fig:mel_spectogram}) and Mel-Frequency Cepstral Coefficients (MFCCs) (\Cref{fig:mfcc}) are commonly used.
These representations emphasize frequency components relevant to human hearing and are widely employed in sound event detection and audio classification tasks.

Several studies have explored \ac{SED} for sports using different audio representations and classification models. Huang et al.
\cite{huang2011a,huang2012} applied \ac{GMM} to MFCC features for SED to enhance visual annotation pipelines. 
Similarly, Yan et al.\cite{yan2014} employed \ac{SVM} for audio-based annotation in tennis.
Baughman et al.~\cite{baughman2019} introduced the first neural network-based approach for detecting tennis hits using a CNN trained on MFCCs and their delta coefficients.
Since audio representations such as spectrograms and MFCCs can be visualized as images (\Cref{fig:bounce_sound}), image-based classification methods like CNNs have proven highly effective. While their model achieved over 90\% accuracy across all classes, the use of 20 ms time frames limited temporal precision, resulting in low onset accuracy.

In the context of spin detection, Yamamoto et al.\cite{yamamoto2020} demonstrated that spin type (topspin, flat, slice) in tennis could be predicted from sound with over 70\% accuracy. 
Their approach involved transforming the audio signal using FFT, compressing the frequency-domain representation via PCA, and classifying the result using an SVM.
However, tennis ball-racket contact durations (5ms\cite{miller2006}) are significantly longer than in table tennis, where impact times are as short as 1.3–1.8ms\cite{kawazoe2003, kawazoe2004}.
This makes fine-grained analysis in table tennis more challenging. Moreover, the audible signal extends beyond the physical contact due to vibrations and room acoustics, as shown in \Cref{fig:bounce_detection}.
These factors introduce a trade-off between temporal and frequency resolution when using FFT-based methods.
Additionally, the acoustic profile is sensitive to racket properties and grip, with Russell et al.~\cite{russell2017} showing that simply holding the racket alters its frequency response.

Specific to table tennis, Zhang et al.\cite{zhang2006} proposed a bounce detection method using energy peak detection. 
For classification between hit and non-hit events, they compared the Mahalanobis distance and a C-Support Vector Classifier using MFCC features.
More recently, Yu et al.\cite{yu2022} introduced a method for estimating bounce locations on the table using a dynamic thresholding approach, refined with frequency-band energy analysis.
However, none of these methods have addressed the problem of distinguishing between acoustically similar events—such as differentiating serves with or without spin, or identifying the racket used—highlighting a key gap addressed by our work.

% http://practicalcryptography.com/miscellaneous/machine-learning/guide-mel-frequency-cepstral-coefficients-mfccs/

\section{Dataset}
\label{sec:dataset}
Due to the noise present in online recordings, we recorded our own dataset.
The different bounce sounds were recorded with the Zoom H4n Pro Handy Recorder, a directional microphone, at 44.1kHz ensuring consistent high-quality data.
The microphone was always oriented towards the player at distances ranging from 50 cm to 2 m.
The bounce sounds were recorded for 10 racket configurations described in \Cref{tab:racket_combinations}.

\begin{table}
    \centering
    \begin{tabular}{|c|c|c|c|} \hline
        \rowcolor{gray!25}\textbf{Blade} & \textbf{Sponge thickness (mm)} & \textbf{Rubber} & \textbf{Id} \\ \hline
        \multirow{2}{*}{Offensive} & 2.1 & Inverted (offensive) & 1\\
        & \cellcolor{gray!25} 1.8 & \cellcolor{gray!25} Inverted (allround) & \cellcolor{gray!25} 2\\ \hline
        \multirow{2}{*}{Defensive} & 2.1 & Inverted (offensive) & 3\\
        & \cellcolor{gray!25} 1.8 & \cellcolor{gray!25} Inverted (allround) & \cellcolor{gray!25} 4\\ \hline
        \multirow{6}{*}{Allrounder} & 1.2 & Long pips & 5\\
        & \cellcolor{gray!25} 0 & \cellcolor{gray!25} Long pips & \cellcolor{gray!25} 6\\
        & 1.2 & Medium pips & 7\\
        & \cellcolor{gray!25} 2.0 & \cellcolor{gray!25} Short pips & \cellcolor{gray!25} 8\\
        & 2.1 & Inverted (Offensive) & 9\\
        & \cellcolor{gray!25} 2.1 & \cellcolor{gray!25} Anti-spin & \cellcolor{gray!25} 10\\ \hline
    \end{tabular}
    \caption{Racket configurations}
    \label{tab:racket_combinations}
\end{table}

We varied the blade, sponge thickness, and rubber type to cover the most common racket configurations.
The audio samples with spin applied on the ball were recorded for serves.
The number of audio samples is detailed in \Cref{tab:dataset}.
Sounds labeled as ‘other’ include acoustically similar non-bounce events (e.g., falling objects such as pens or tennis balls), added to improve classifier robustness.
The onset of the bounce was manually annotated at the beginning of the bounce's waveform.
We split the dataset into 80\% training and 20\% testing sets, ensuring that the distribution of racket types and spin labels was preserved across both partitions.

\begin{table}
    \centering
    \begin{tabular}{|c|ccc|c|}
        \hline
        \rowcolor{gray!25} \textbf{Surface} & \textbf{Back} & \textbf{Flat} & \textbf{Top} & \textbf{Total}\\
        \hline
        Racket 01 & 263 & 354 & 275 & 892 \\
        \rowcolor{gray!25}Racket  02 & 93 & 168 & 40 & 301 \\
        Racket 03 & 70 & 162 & 43 & 275 \\
        \rowcolor{gray!25}Racket 04 & 98 & 145 & 45 & 288 \\
        Racket 05 & 55 & 185 & 0 & 240 \\
        \rowcolor{gray!25} Racket 06 & 55 & 152 & 0 & 207 \\
        Racket 07 & 60 & 184 & 0 & 244 \\
        \rowcolor{gray!25} Racket 08 & 101 & 193 & 42 & 336 \\
        Racket 09 & 96 & 159 & 41 & 296 \\
        \rowcolor{gray!25} Racket 10 & 100 & 177 & 40 & 317 \\
        \hline
        \textbf{Total} & 991 & 1879 & 526 & 3396 \\
        \hline
        \rowcolor{gray!25}Table & \multicolumn{4}{c|}{777} \\
        Floor & \multicolumn{4}{c|}{290} \\
        \rowcolor{gray!25}Other & \multicolumn{4}{c|}{1239} \\
        \hline

    \end{tabular}
    \caption{Distribution of samples across different classes in the dataset. The dataset was subsequently divided into training and testing subsets with an 80/20 split ratio.}
    \label{tab:dataset}
\end{table}

\begin{figure*}
    \centering
    \begin{subfigure}[b]{0.28\linewidth}
        \centering
        \includegraphics[width=\linewidth]{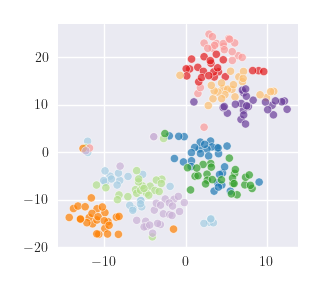}
        \caption{Simple bounces}
        \label{fig:tsne_few}
    \end{subfigure}
    \hfill
    \begin{subfigure}[b]{0.32\linewidth}
        \centering
        \includegraphics[width=\linewidth]{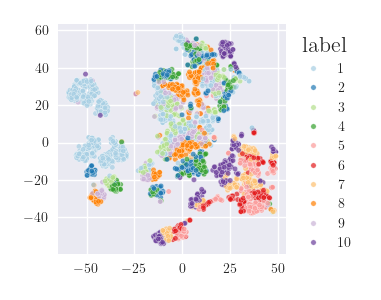}
        \caption{All bounces}
        \label{fig:tsne_all}
    \end{subfigure}
    \hfill
    \begin{subfigure}[b]{0.36\linewidth}
        \centering
        \includegraphics[width=\linewidth]{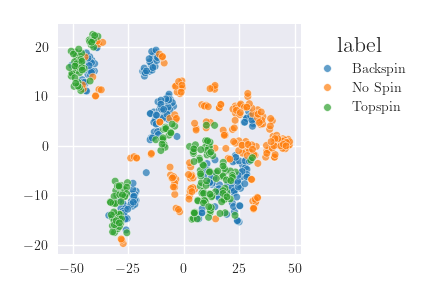}
        \caption{Spin classification}
        \label{fig:tsne_spin}
    \end{subfigure}
    \caption{
t-SNE visualizations of bounce sound embeddings derived from Mel spectrograms. In (a) and (b), we show racket type clustering, while in (c) we show spin clustering.
(a) Clustering of sounds from vertical racket bounces reveals that racket types produce distinguishable acoustic signatures. 
(b) Including all strokes—including serves with and without spin—introduces more variability, yet distinct clusters remain visible, indicating robustness to stroke type. 
(c) Clustering based on spin shows that the presence of spin can be inferred from the sound, as spin and no-spin bounces form separable groups.
}
    \label{fig:tsne_combined}
\end{figure*}

As a sanity check to determine if the sounds are distinguishable, we applied t-SNE to the normalized Mel spectrograms of different bounce samples and plotted their 2D representation in \Cref{fig:tsne_combined}.
For very simple vertical racket bounces in \Cref{fig:tsne_few}, we notice clear clusters for each racket.
Rackets 1/3 and 2/4 are confused together.
This seems to indicate that rubber type has more influence on the sound than the blade type.
In \Cref{fig:tsne_all}, we also include racket-ball sounds generated during serves, both with and without spin.
While the sounds still exhibit noticeable clustering, the separation is less distinct, highlighting the greater complexity and variability of real-game bounces, which can lead to overlapping acoustic signatures.

\begin{figure}
    \centering
    \includegraphics[width=0.9\linewidth]{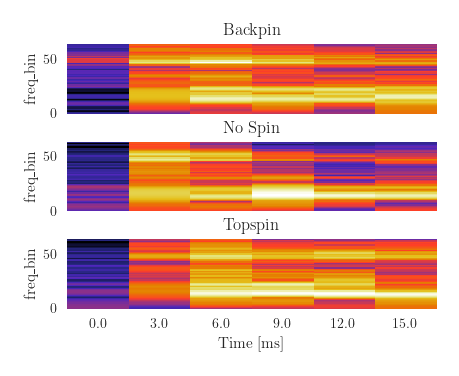}
    \caption{Spectrograms of ball bounces against racket 10 with different spin types. Topspin and backspin bounces show stronger high-frequency components compared to no-spin, indicating that spin affects the acoustic signature of the impact.}
    \label{fig:spectogram}
\end{figure}

For the spin on \Cref{fig:tsne_spin}, we notice that backspin and topspin are often together in clusters.
We can further observe in \Cref{fig:spectogram} that the backspin and topspin spectrograms, though different from no spin spectogram, are almost identical. 
This aligns with the findings from~\cite{park2016, peterossi2017}, which suggest that while the magnitude of spin can be discerned from the sound, identifying the spin direction is more challenging.

\section{Bounce detection}

Sound has been shown to improve table tennis players' response times and counter-attacks, as the bounce sound provides a highly accurate timing cue.
Therefore, we want our bounce detector to achieve the highest possible temporal accuracy. In \ac{SED}, the goal is to detect both the onset and offset of the event.
However, given that ball bounces are extremely brief, we treat the event as discrete and disregard the offset.

Most \ac{SED} methods extract time frames from the audio signal and perform classification on these frames~\cite{mesaros2021}.
However, this approach restricts temporal accuracy to the size of the time frames.
Although reducing the frame size could improve accuracy, it would also result in a loss of the frequency resolution.
Instead, we adopt a two-stage approach described in \Cref{fig:pipeline}, similar to \cite{zhang2006, yu2022}.
First, we use energy-based peak detection to identify potential bounce sounds.
Then, a CNN classifies the detected peaks.

\begin{figure}
    \centering
    \includegraphics[width=\linewidth]{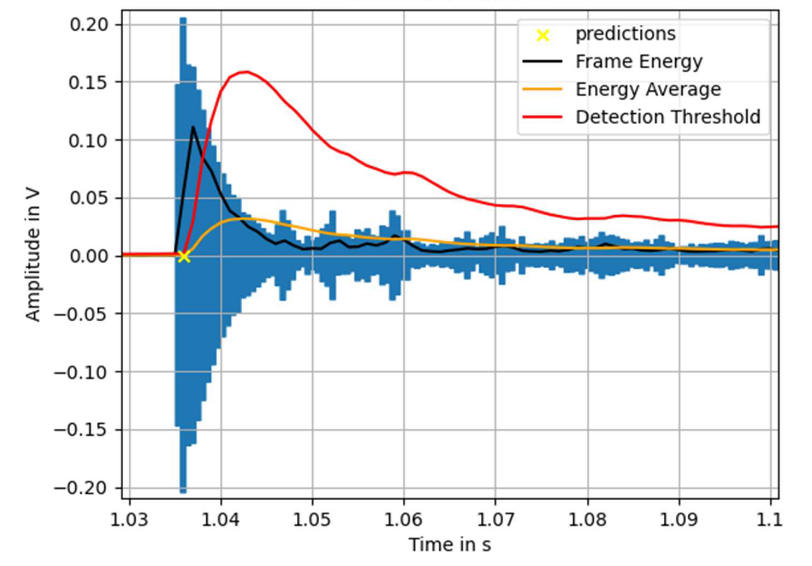}
    \caption{Bounce detection using energy peak tracking. The frame-level energy (black) is monitored against a dynamic threshold (red), computed as an exponential moving average of past energy values (orange). A bounce event is detected when the instantaneous energy sharply exceeds this adaptive threshold.}
    \label{fig:bounce_detection}
\end{figure}

In \Cref{fig:bounce_detection}, we show the evolution of the audio signal for a bounce sound.
The frame energy $e$ is the mean energy across a concise time frame, in our case 1 ms.
Instead of comparing that frame energy to other frame's energies, we compare it to a moving decaying average.
This exponential moving average,

\begin{equation}
    \overline{E}_{k+1} = \gamma \cdot\overline{E}_k + (1-\gamma)\cdot e,
    \label{eq:moving_avg}
\end{equation}

where $\gamma$ is the decay factor for the moving average, allows real-time detection compared to~\cite{zhang2006}.

A multiplier is applied to the moving average to set the detection threshold, chosen empirically for optimal recall and accuracy.
Whenever the frame energy peaks above this detection threshold, a peak is recorded.
This allows for energy peak detection with minimal latency, as sudden energy peaks generally get detected within a millisecond of their onset.
While this approach works effectively in a noiseless environment, other sounds can degrade the detector's performance.
Therefore, the audio signal is first filtered using a 5th-order Butterworth high-pass filter with a cutoff frequency of 10 kHz.
We apply zero-phase filtering to ensure high temporal accuracy.
This filter is chosen because table tennis ball bounces consistently exhibit energy around 11 kHz, unlike human speech.

\section{Classification}

Although the bounce detection provides high onset temporal accuracy, it may detect peaks generated by sources other than ball bounces such as a falling pen.
To refine the results, we use a CNN-based classifier.

\subsection{Input and feature extraction}

The preprocessing steps for the classifier involve loading and preparing the training data, ensuring the audio is sampled at 44.1 kHz, and standardizing the length to 661 samples per data point, corresponding to 15 ms of audio.
This duration was selected based on the typical temporal profile of bounce sounds, as illustrated in \Cref{fig:bounce_detection}, where the main acoustic signature of a bounce is concentrated within this window.
The stereo audio is converted to mono and then transformed into a Mel spectrogram. 
Experimental results demonstrated that the CNN performed best using 64 Mel bands and an FFT window length of 256 samples. 
We also compared different input representations, including raw spectrograms, MFCCs, and LFCCs.
The Mel spectrogram consistently produced the best results.

\subsection{Model}

\begin{table}
\centering
\begin{tabular}{|c|c|c|c|}
\hline
\textbf{Layer} & \textbf{Type} & \textbf{Kernel Size} & \textbf{Stride} \\ \hline
1 & Conv2d + ReLU + BatchNorm2d & 5 & (2, 2) \\ \hline
2 & Conv2d + ReLU + BatchNorm2d & 3 & (2, 1) \\ \hline
3 & Conv2d + ReLU + BatchNorm2d & 3 & (2, 1) \\ \hline
4 & Conv2d + ReLU + BatchNorm2d & 3 & (2, 1) \\ \hline
5 & Conv2d + ReLU + BatchNorm2d & 3 & (2, 1) \\ \hline
6 & Conv2d + ReLU + BatchNorm2d & 3 & (2, 1) \\ \hline
7 & AdaptiveAvgPool2d & Output Size = 1 & - \\ \hline
8 & Linear & - & - \\ \hline
9 & Softmax & - & - \\ \hline
\end{tabular}
\caption{Neural Network Architecture}
\label{tab:nn_architecture}
\end{table}

We employ a six-layer \ac{CNN} to classify bounce sounds, where each layer consists of a 2D convolution followed by a ReLU activation and batch normalization, as detailed in \Cref{tab:nn_architecture}.
The model is trained using the categorical cross-entropy loss.
To address the two classification tasks—surface type (different rackets, table, floor, other) and spin type (backspin, topspin, no spin), we train two separate instances of the same network architecture, allowing each model to specialize in its respective task.

\section{Results}

\subsection{Peak detection}

The energy peak detection achieves solid results on test data, shown in \Cref{tab:detector-results}.
These results remain consistent even with human speech, recorded in the same condition, being laid over the data, simulating a noisy environment.
The detection algorithm shows high temporal resolution, detecting energy peaks on average no later than 0.2 ms after occurring.
The increased delay when there is background noise is due to the peak being proportionally less pronounced.

\begin{table}
\centering
\begin{tabular}{|c|c|c|c|}
\hline
 & \textbf{Precision} & \textbf{Recall} & \textbf{Onset accuracy} (ms)\\ \hline
No Noise & 0.98 & 1 & 0.09\\
Noise & 0.98 & 0.95 & 0.2\\ \hline
\end{tabular}
\caption{Peak detection results}
\label{tab:detector-results}
\end{table}

\subsection{Bounce classification}

\begin{figure}
    \centering
    \includegraphics[width=0.6\linewidth]{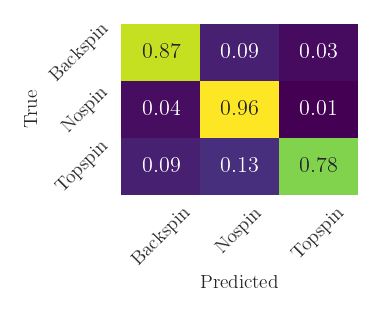}
    \caption{Confusion matrices for spin classification}
    \label{fig:cm-spin}
\end{figure}

\begin{figure}
    \centering
    \includegraphics[width=\linewidth]{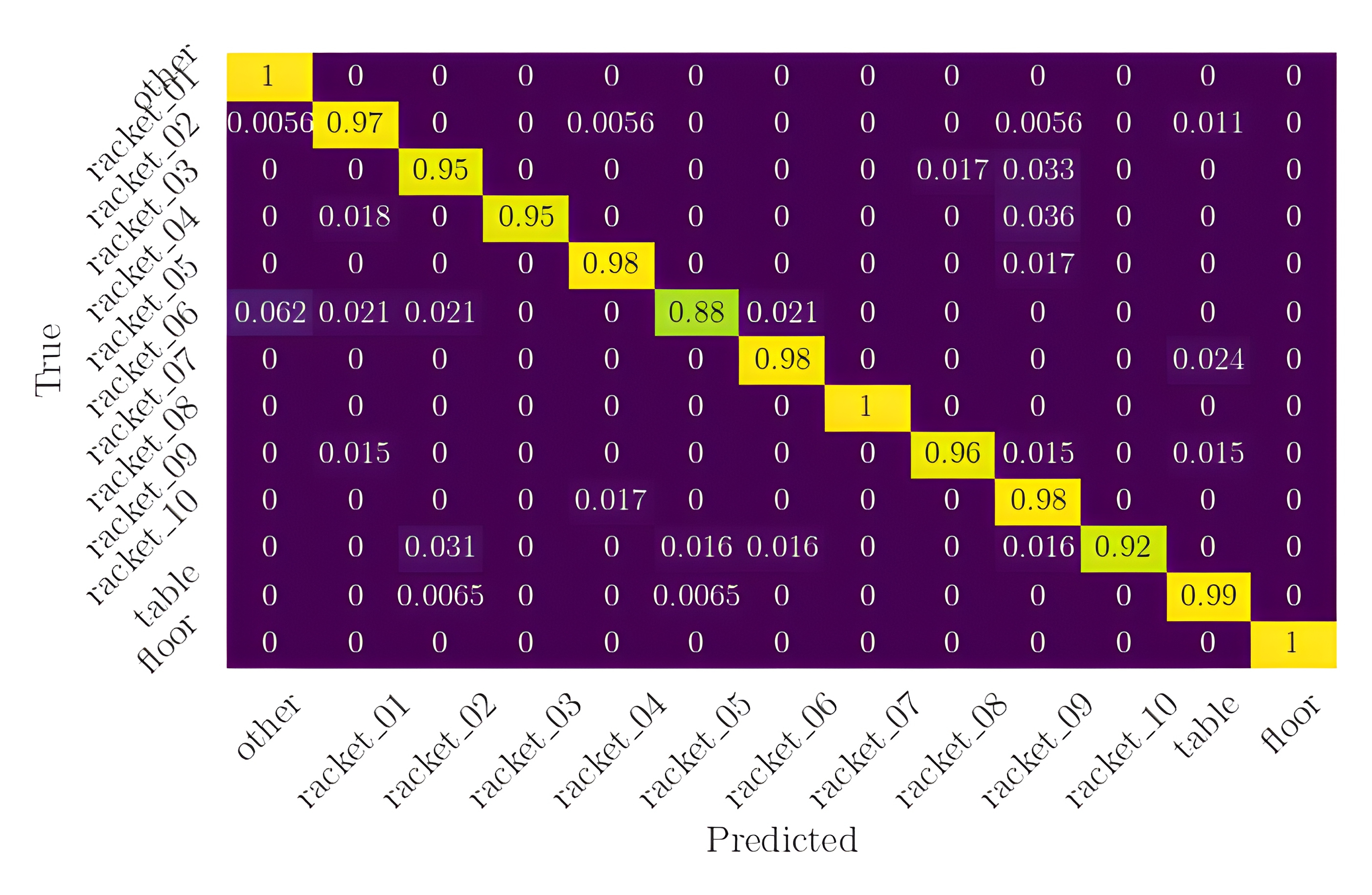}
    \caption{Confusion matrix for bounce surface classification}
    \label{fig:cm_surf}
\end{figure}

To demonstrate the feasibility of deploying our classification pipeline on edge devices for real-time use in gymnasiums or training environments, we benchmarked the inference time of our CNN model on a \textit{Core i7-1165G7 2.8\,GHz $\times$8 CPU}.  
The model achieves an average inference time of 4.2\,ms $\pm$ 1.1\,ms, confirming its suitability for low-latency, real-time applications.

We evaluated the performance of our CNN classifier against two traditional approaches commonly used in sound event detection for table tennis: a \ac{GMM} with diagonal covariance matrices~\cite{huang2011a,huang2012}, and an \ac{SVM} with a linear kernel~\cite{yamamoto2020}.  

\Cref{fig:cm-spin} shows the confusion matrix for the spin classification task, with accuracies of 87\% for \textit{backspin}, 96\% for \textit{no-spin}, and 78\% for \textit{topspin}.  
The main confusions occur between \textit{topspin} and \textit{backspin}, reflecting their acoustic similarity.  
Some spin samples are misclassified as \textit{no-spin}, likely due to weak spin application that produces insufficient acoustic cues despite being labeled as spin.  
Overall, the model effectively detects spin presence, though direction classification and borderline cases remain challenging.

As shown in \Cref{tab:classification_evaluation}, the CNN outperforms both baselines across all metrics, with the \ac{SVM} coming close in performance.  
The \ac{GMM} performs poorly, likely due to its limited capacity to distinguish between acoustically similar bounce sounds using simplistic distributional assumptions.

The surface classification model reliably identifies the bounce surface as table, racket, floor, or other, as illustrated in the confusion matrix in \Cref{fig:cm_surf}.  
However, finer-grained distinctions—particularly between rackets—prove more challenging. 
Racket 6, which lacks a sponge layer, is the hardest to classify accurately.  
Most classification errors occur between rackets with similar sponge thicknesses, whereas differences in blade type yield sufficiently distinct acoustic signatures to be discriminated even when using the same rubber.

The dataset was recorded under controlled conditions with a single player, which may limit generalization to unseen players, rackets, or environments.
Moreover, distinguishing spin directions remains difficult due to their acoustic similarity, and bounce sounds can be influenced by player-specific technique.
Future work should therefore evaluate generalization across racket types, playing styles, and recording conditions.

\begin{table}
    \centering
    \begin{tabular}{|c|c|c|c|c|c|c|}
        \hline
        \textbf{Method} & \multicolumn{2}{|c|}{\textbf{Precision}} & \multicolumn{2}{|c|}{\textbf{Recall}} & \multicolumn{2}{|c|}{\textbf{F1}} \\
        \hline
        Task & Spin & Surf. & Spin & Surf. & Spin & Surf.\\
        \hline
        \ac{GMM} & 0.22 & 0.09 & 0.25 & 0.06 & 0.17 & 0.06\\
        \ac{SVM} &0.96 & 0.94 & 0.92 & 0.93 & 0.93 & 0.94 \\
        CNN & \textbf{0.98} & \textbf{0.97} & \textbf{0.95} & \textbf{0.97} & \textbf{0.96} & \textbf{0.97} \\
        \hline
    \end{tabular}
    \caption{Evaluation of the different methods for the classification}
    \label{tab:classification_evaluation}
\end{table}

\section{Conclusion}
In this paper, we demonstrated that auditory information from table tennis bounce sounds contains valuable cues about the ball’s spin and the type of racket used.
We proposed a real-time-capable pipeline combining high-precision bounce detection with CNN-based classification, trained on a comprehensive dataset of labeled bounce sounds.
Our system accurately distinguishes between bounce surfaces and reliably identifies whether spin was applied to the ball.

Beyond robotics, this approach opens new possibilities for sports coaching and performance analysis. 
By delivering immediate acoustic feedback, the system can support players and coaches in evaluating shot quality, spin application, and equipment characteristics—especially in situations where visual cues are limited or delayed, such as in game recordings where player occlusion frequently occurs.
The method also lays the groundwork for non-invasive studies of player technique and material properties.

Future work may explore using sound to estimate specific racket attributes such as blade stiffness, sponge thickness, or rubber type, further enhancing the understanding and optimization of player–equipment interaction.

% \cite{pham2018} Object detector could be used to detect the bounce sound, running object detection algorithms such as faster RCNN on the Mel-spectogram.

% \cite{shao2023} Fine-tuning

\section*{Acknowledgment}
We would like to thank Jonas Tebbe for sharing his table tennis experience and his help for the serve recordings.

\bibliographystyle{IEEEtran}
\bibliography{IEEEabrv, bilbio}

\begin{thebibliography}{10}
\providecommand{\url}[1]{#1}
\csname url@samestyle\endcsname
\providecommand{\newblock}{\relax}
\providecommand{\bibinfo}[2]{#2}
\providecommand{\BIBentrySTDinterwordspacing}{\spaceskip=0pt\relax}
\providecommand{\BIBentryALTinterwordstretchfactor}{4}
\providecommand{\BIBentryALTinterwordspacing}{\spaceskip=\fontdimen2\font plus
\BIBentryALTinterwordstretchfactor\fontdimen3\font minus \fontdimen4\font\relax}
\providecommand{\BIBforeignlanguage}[2]{{%
\expandafter\ifx\csname l@#1\endcsname\relax
\typeout{** WARNING: IEEEtran.bst: No hyphenation pattern has been}%
\typeout{** loaded for the language `#1'. Using the pattern for}%
\typeout{** the default language instead.}%
\else
\language=\csname l@#1\endcsname
\fi
#2}}
\providecommand{\BIBdecl}{\relax}
\BIBdecl

\bibitem{canal-bruland2018}
R.~{Ca{\~n}al-Bruland}, F.~M{\"u}ller, B.~Lach, and C.~Spence, ``Auditory contributions to visual anticipation in tennis,'' \emph{Psychology of Sport and Exercise}, vol.~36, pp. 100--103, May 2018.

\bibitem{canal-bruland2022}
R.~{Ca{\~n}al-Bruland}, H.~S. Meyerhoff, and F.~M{\"u}ller, ``Context modulates the impact of auditory information on visual anticipation,'' \emph{Cognitive Research: Principles and Implications}, vol.~7, p.~76, Aug. 2022.

\bibitem{sors2017}
F.~Sors, M.~Murgia, I.~Santoro, V.~Prpic, A.~Galmonte, and T.~Agostini, ``The contribution of early auditory and visual information to~the~discrimination of shot power in ball sports,'' \emph{Psychology of Sport and Exercise}, vol.~31, pp. 44--51, Jul. 2017.

\bibitem{takeuchi1993}
T.~Takeuchi, ``Auditory information in playing tennis,'' \emph{Perceptual and Motor Skills}, vol.~76, no. 3 Pt 2, pp. 1323--1328, Jun. 1993.

\bibitem{fujita2023a}
R.~A. Fujita, D.~P.~R. Santos, R.~N. Barbosa, L.~H. Palucci~Vieira, P.~R.~P. Santiago, A.~M. Zagatto, and M.~M. Gomes, ``Auditory {{Information Reduces Response Time}} for {{Ball Rotation Perception}}, {{Increasing Counterattack Performance}} in {{Table Tennis}},'' \emph{Research Quarterly for Exercise and Sport}, vol.~94, no.~1, pp. 55--63, Mar. 2023.

\bibitem{park2016}
S.~H. Park, S.~Kim, M.~Kwon, and E.~A. Christou, ``Differential contribution of visual and auditory information to accurately predict the direction and rotational motion of a visual stimulus,'' \emph{Applied Physiology, Nutrition, and Metabolism = Physiologie Appliquee, Nutrition Et Metabolisme}, vol.~41, no.~3, pp. 244--248, Mar. 2016.

\bibitem{peterossi2017}
D.~Peterossi, R.~Negri~Barbosa, L.~Palucci~Vieira, P.~Santiago, A.~Zagatto, and M.~Gomes, ``Training {{Level Does Not Affect Auditory Perception}} of {{The Magnitude}} of {{Ball Spin}} in {{Table Tennis}},'' \emph{Journal of Human Kinetics}, vol.~55, pp. 19--27, Jan. 2017.

\bibitem{petri2020}
K.~Petri, T.~Schmidt, and K.~Witte, ``The influence of auditory information on performance in table tennis,'' \emph{European Journal of Human Movement}, vol.~45, Dec. 2020.

\bibitem{wang2023}
Y.~Wang, Z.~Sun, Y.~Luo, H.~Zhang, W.~Zhang, K.~Dong, Q.~He, Q.~Zhang, E.~Cheng, and B.~Song, ``A {{Novel Trajectory-Based Ball Spin Estimation Method}} for {{Table Tennis Robot}},'' \emph{IEEE Transactions on Industrial Electronics}, pp. 1--11, 2023.

\bibitem{su2013}
H.~Su, Z.~Fang, D.~Xu, and M.~Tan, ``Trajectory {{Prediction}} of {{Spinning Ball Based}} on {{Fuzzy Filtering}} and {{Local Modeling}} for {{Robotic Ping}}--{{Pong Player}},'' \emph{Instrumentation and Measurement, IEEE Transactions on}, vol.~62, pp. 2890--2900, Nov. 2013.

\bibitem{tebbe2020a}
J.~Tebbe, L.~Klamt, Y.~Gao, and A.~Zell, ``Spin {{Detection}} in {{Robotic Table Tennis}},'' in \emph{2020 {{IEEE International Conference}} on {{Robotics}} and {{Automation}} ({{ICRA}})}, May 2020, pp. 9694--9700.

\bibitem{chen2010a}
X.~Chen, Y.~Tian, Q.~Huang, W.~Zhang, and Z.~Yu, ``Dynamic model based ball trajectory prediction for a robot ping-pong player,'' \emph{2010 IEEE International Conference on Robotics and Biomimetics, ROBIO 2010}, Dec. 2010.

\bibitem{sato2020}
S.~Sato and M.~Aono, ``Leveraging {{Human Pose Estimation Model}} for {{Stroke Classification}} in {{Table Tennis}},'' \emph{MediaEval'20}, 2020.

\bibitem{kulkarni2021}
K.~M. Kulkarni and S.~Shenoy, ``Table {{Tennis Stroke Recognition Using Two-Dimensional Human Pose Estimation}},'' May 2021.

\bibitem{gao2021}
Y.~Gao, J.~Tebbe, and A.~Zell, ``Robust {{Stroke Recognition}} via {{Vision}} and~{{IMU}} in {{Robotic Table Tennis}},'' in \emph{Artificial {{Neural Networks}} and {{Machine Learning}} -- {{ICANN}} 2021}, ser. Lecture {{Notes}} in {{Computer Science}}, I.~Farka{\v s}, P.~Masulli, S.~Otte, and S.~Wermter, Eds.\hskip 1em plus 0.5em minus 0.4em\relax Cham: Springer International Publishing, 2021, pp. 379--390.

\bibitem{gossard2024}
T.~Gossard, J.~Krismer, A.~Ziegler, and A.~Zell, ``Table tennis ball spin estimation with an event camera,'' \emph{Proceedings of the IEEE/CVF Conference on Computer Vision and Pattern Recognition}, pp. 3347--3356, 2024.

\bibitem{gossard2023}
T.~Gossard, J.~Tebbe, A.~Ziegler, and A.~Zell, \emph{{{SpinDOE}}: {{A Ball Spin Estimation Method}} for {{Table Tennis Robot}}}.\hskip 1em plus 0.5em minus 0.4em\relax IEEE, Oct. 2023.

\bibitem{xiong2003}
Z.~Xiong, R.~Radhakrishnan, A.~Divakaran, and T.~Huang, \emph{Audio Events Detection Based Highlights Extraction from Baseball, Golf and Soccer Games in a Unified Framework}.\hskip 1em plus 0.5em minus 0.4em\relax IEEE, Aug. 2003, vol.~5.

\bibitem{baillie2003}
M.~Baillie and J.~Jose, \emph{Audio-{{Based Event Detection}} for {{Sports Video}}}.\hskip 1em plus 0.5em minus 0.4em\relax Springer Berlin Heidelberg, Jul. 2003, vol. 2728.

\bibitem{huang2011a}
\BIBentryALTinterwordspacing
Q.~Huang, S.~J. Cox, F.~Yan, T.~E. de~Campos, D.~Windridge, J.~Kittler, and W.~J. Christmas, ``Improved detection of ball hit events in a tennis game using multimodal information,'' in \emph{AVSP ..}, 2011. [Online]. Available: \url{https://api.semanticscholar.org/CorpusID:15687336}
\BIBentrySTDinterwordspacing

\bibitem{huang2012}
Q.~Huang and S.~Cox, \emph{Improved Audio Event Detection by Use of Contextual Noise}.\hskip 1em plus 0.5em minus 0.4em\relax IEEE, Mar. 2012.

\bibitem{yan2014}
F.~Yan, J.~Kittler, D.~Windridge, W.~Christmas, K.~Mikolajczyk, S.~Cox, and Q.~Huang, ``Automatic annotation of tennis games: {{An}} integration of audio, vision, and learning,'' \emph{Image and Vision Computing}, vol.~32, no.~11, pp. 896--903, Nov. 2014.

\bibitem{baughman2019}
A.~Baughman, E.~Morales, G.~Reiss, N.~Greco, S.~Hammer, and S.~Wang, ``Detection of {{Tennis Events}} from {{Acoustic Data}},'' in \emph{Proceedings {{Proceedings}} of the 2nd {{International Workshop}} on {{Multimedia Content Analysis}} in {{Sports}}}.\hskip 1em plus 0.5em minus 0.4em\relax Nice France: ACM, Oct. 2019, pp. 91--99.

\bibitem{yamamoto2020}
N.~Yamamoto, K.~Nishida, K.~Itoyama, and K.~Nakadai, \emph{Detection of {{Ball Spin Direction}} Using {{Hitting Sound}} in {{Tennis}}}.\hskip 1em plus 0.5em minus 0.4em\relax SCITEPRESS - Science and Technology Publications, Jan. 2020.

\bibitem{miller2006}
S.~Miller, ``Modern tennis rackets, balls, and surfaces,'' \emph{British Journal of Sports Medicine}, vol.~40, no.~5, pp. 401--405, May 2006.

\bibitem{kawazoe2003}
Y.~Kawazoe and D.~SUZUKI, ``Prediction of {{Table Tennis Racket Restitution Performance Based}} on the {{Impact Analysis}},'' \emph{Theoretical and Applied Mechanics Japan}, vol.~52, pp. 163--174, Jan. 2003.

\bibitem{kawazoe2004}
Y.~Kawazoe and D.~Suzuki, \emph{Impact Prediction between Ball and Racket in Table Tennis, {{Science}} and {{Racket Sports III}}, {{Routledge}}, Pp.134- 139 (2004).}, Jan. 2004.

\bibitem{russell2017}
D.~Russell, ``Vibroacoustic analysis of table tennis rackets and balls: {{The}} acoustics of ping pong,'' \emph{The Journal of the Acoustical Society of America}, vol. 141, pp. 3979--3979, May 2017.

\bibitem{zhang2006}
B.~Zhang, W.~Dou, and L.~Chen, \emph{Ball {{Hit Detection}} in {{Table Tennis Games Based}} on {{Audio Analysis}}}.\hskip 1em plus 0.5em minus 0.4em\relax IEEE, Jan. 2006, vol.~3.

\bibitem{yu2022}
H.-I. Yu, S.-C. Hong, and T.-Y. Ju, ``Low-cost system for real-time detection of the ball-table impact position on ping-pong table,'' \emph{Applied Acoustics}, vol. 195, p. 108832, Jun. 2022.

\bibitem{mesaros2021}
A.~Mesaros, T.~Heittola, T.~Virtanen, and M.~D. Plumbley, ``Sound {{Event Detection}}: {{A Tutorial}},'' \emph{IEEE Signal Processing Magazine}, vol.~38, no.~5, pp. 67--83, Sep. 2021.

\end{thebibliography}

\end{document}